\documentclass[preprint,amsmath,amssymb,aps,showkeys,showpacs]{revtex4}
\usepackage[english]{babel}
\usepackage{graphicx}
\usepackage{graphics}
\usepackage{amsmath}
\usepackage{dcolumn}
\usepackage{amssymb}
\usepackage{bm}

\begin{document}
\title{Aharonov-Bohm-type effect in the background of a distortion of a vertical line into a vertical spiral}
\author{W. C. F. da Silva}
\affiliation{Departamento de F\'isica, Universidade Federal da Para\'iba, Caixa Postal 5008, 58051-900, Jo\~ao Pessoa, PB, Brazil.}

\author{K. Bakke}
\email{kbakke@fisica.ufpb.br}
\affiliation{Departamento de F\'isica, Universidade Federal da Para\'iba, Caixa Postal 5008, 58051-900, Jo\~ao Pessoa, PB, Brazil.}

\begin{abstract}

It is analysed Aharonov-Bohm-type effects when a spinless quantum particle is in an elastic medium with the distortion of a vertical line into a vertical spiral. By confining the spinless particle to a cylindrical box, the analogue of the Aharonov-Bohm effect for bound states is observed due to influence of the topological defect on the allowed energies of the system. It corresponds to the shift in the angular momentum quantum number yielded by the effects of the topology of the distortion of a vertical line into a vertical spiral. In addition, it is analysed the effects of rotation. It is shown that the Aharonov-Bohm effect for bound states exists. Besides, there is an analogue of the coupling between the angular momentum and angular velocity of the rotating frame.
 
\end{abstract}


\maketitle

\section{Introduction}

Based on the elastic theory in solids, dislocations arise from a break of the translational symmetry \cite{kleinert,kat}. Besides, it characterizes the presence of torsion in the elastic medium. At present days, it is well-known that dislocations can be described by the differential geometry, i.e., the deformation of the elastic medium that corresponds to a dislocation can be described by the Riemann-Cartan geometry. This proposal of using the Riemann-Cartan geometry was proposed by Katanaev and Volovich \cite{kat}, and hence, the process of ``cut'' and ``glue'' of the elastic medium (Volterra process \cite{kleinert}) was connected with the geometrical description of the a topological defect in the context of general relativity. Dislocations can modify the electronic properties of the medium, therefore, there is a great interest in studies of semiconductors \cite{semi,semi3,semi4} and quantum dots \cite{au}. This kind of elastic deformation in a solid is called as a linear topological defect. Examples of it are the screw dislocation and the spiral dislocation \cite{val,put}. In recent decades, the Katanaev-Volovich approach \cite{kat} has been widely used in the literature with the purpose of investigating topological effects associated with screw dislocations in quantum rings \cite{fur6,fur3,dantas1}, electrons subject to the deformed Kratzer potential \cite{1}, electron gas in a cylindrical shell \cite{shell}, quantum holonomies \cite{bf3} and with an electron in a uniform magnetic field \cite{fur,fur2,fur7,fil}. Furthermore, topological effects associated with a spiral dislocation have been made with geometric quantum phases \cite{bf} and the harmonic oscillator \cite{mb}.

In this work, we explore this line of research and investigate the topological effects associated with a screw dislocation on a spinless particle confined to a hard-wall confining potential. The most known kind of screw dislocation in the literature is the distortion of a circular curve into a vertical spiral. As shown in Refs. \cite{fur2,fur6}, the main effect of the presence of the distortion of a circular curve into a vertical spiral is the shift in the angular momentum quantum number that gives rise to an analogue effect of the Aharonov-Bohm effect for bound states \cite{pesk,ab}. With the interface between general relativity and relativistic quantum mechanics, spacetime with dislocations have drawn attention to relativistic analogue effects of the Aharonov-Bohm effect \cite{valdir,8,31,32}. In the present work, we deal with another type of screw dislocation that corresponds to the distortion of a vertical line into a vertical spiral \cite{val}. Then, we analyse Aharonov-Bohm-type effects that stem from the presence of the distortion of a vertical line into a vertical spiral in the elastic medium. In addition, we investigate effects of rotation on this system.

This paper is structured as follows: in section II, we introduce the line element of the distortion of a vertical line into a vertical spiral. Then, we obtain the Schr\"odinger equation in the presence of this topological defect and analyse the confinement of a spinless quantum particle to a hard-wall confining potential; in section III, we investigate effects of rotation; in section IV, we present our conclusions.

\section{Aharonov-Bohm-type effect}

In this section, we search for analytical solutions to the Schr\"odinger equation for a spinless particle in the presence of a screw dislocation. Here, we consider a screw dislocation that corresponds to the distortion of a vertical line into a vertical spiral \cite{put,val}. This kind of topological defect is described by the line element:
\begin{eqnarray}
ds^{2}=dr^{2}+r^{2}d\varphi^{2}+2\beta\,d\varphi\,dz+dz^{2}
\label{1.1}
\end{eqnarray}
where $0\,<\,r\,<\,\infty$, $0\leq\varphi\leq2\pi$ and $-\infty\,<\,z\,<\,\infty$ and we have used the units with $c=1$ and $\hbar=1$. The parameter $\beta$ is a constant that characterizes the torsion field (dislocation). It can be defined in the range $0\,<\,\beta\,<\,1$.

According to Refs.  \cite{fur,fur2,fur7}, the time-independent Schr\"odinger equation in the presence of a topological defect is written in the form (with the units: $c=1$ and $\hbar=1$):
\begin{eqnarray}
\mathcal{E}\psi=-\frac{1}{2m}\frac{1}{\sqrt{g}}\,\partial_{i}\left(g^{ij}\,\sqrt{g}\,\partial_{j}\right)\psi,
\label{1.1a}
\end{eqnarray}
where $g_{ij}$ is the metric tensor, $g^{ij}$ is the inverse of $g_{ij}$ and $g=\mathrm{det}\left|g_{ij}\right|$. Observe that the indices $\left\{i,\,j\right\}$ run over the space coordinates. Thereby, with the line element (\ref{1.1}), the time-independent Schr\"odinger equation (\ref{1.1a}) becomes
\begin{eqnarray}
-2m\mathcal{E}\psi=\frac{\partial^{2}\psi}{\partial r^{2}}+\frac{r}{\left(r^{2}-\beta^{2}\right)}\,\frac{\partial\psi}{\partial r}+\frac{1}{\left(r^{2}-\beta^{2}\right)}\left[\frac{\partial}{\partial\varphi}-\beta\frac{\partial}{\partial z}\right]^{2}\psi+\frac{\partial^{2}\psi}{\partial z^{2}}.
\label{1.2}
\end{eqnarray}

It is worth observing that the cylindrical symmetry is present in the line element (\ref{1.1}) and, as a consequence, in the Schr\"odinger equation (\ref{1.2}). In this way, a possible way of writing the solution to Eq. (\ref{1.2}) is $\psi\left(r,\,\varphi,\,z\right)=e^{il\varphi+ikz}\,R\left(r\right)$, where $k=\mathrm{const}$ and $l=0,\pm1,\pm2,\pm3\ldots$ are the eigenvalues of the operators $\hat{p}_{z}=-i\partial_{z}$ and $\hat{L}_{z}=-i\partial_{\varphi}$, respectively. Thereby, from  Eq. (\ref{1.2}), we obtain a radial equation given in the form:
\begin{eqnarray}
R''+\frac{r}{\left(r^{2}-\beta^{2}\right)}\,R'-\frac{\left(l-\beta\,k\right)^{2}}{\left(r^{2}-\beta^{2}\right)}\,R+\left(2m\mathcal{E}-k^{2}\right)\,R=0.
\label{1.3}
\end{eqnarray}

Let us proceed our discussion by defining:
\begin{eqnarray}
r=\beta\,\cosh y.
\label{1.4}
\end{eqnarray}
Then, the radial equation (\ref{1.3}) becomes
\begin{eqnarray}
R''+\left[2\,q^{2}\,\cosh\left(2y\right)-\lambda\right]R=0,
\label{1.5}
\end{eqnarray}
where we have defined the following parameters in Eq. (\ref{1.5}):
\begin{eqnarray}
\lambda&=&\left(l-\beta\,k\right)^{2}+\frac{\beta^{2}}{2}\left(2m\mathcal{E}-k^{2}\right);\nonumber\\
[-2mm]\label{1.6}\\[-2mm]
q^{2}&=&\frac{\beta^{2}}{2}\left(2m\mathcal{E}-k^{2}\right).\nonumber
\end{eqnarray}
Hence, Eq. (\ref{1.5}) is known in the literature as the modified Mathieu equation or radial Mathieu equation \cite{abra,arf,mat1,mat2}. 

With the purpose of investigating the confinement of a spinless quantum particle to a hard-wall confining potential, let us define the parameter \cite{mat2}:
\begin{eqnarray}
x&=&\beta\sqrt{2m\mathcal{E}-k^{2}}\,\cosh y\nonumber\\
&=&2q\,\cosh y.
\label{1.8}
\end{eqnarray}
Then, by substituting Eq. (\ref{1.8}) into Eq. (\ref{1.5}), we obtain a second-order differential equation given by:
\begin{eqnarray}
R''+\frac{1}{x}\,R'-\frac{1}{x^{2}}\left[\lambda\,R+2q^{2}\left(R+2R''\right)\right]+R=0.
\label{1.9}
\end{eqnarray}

Observe that the parameter that characterizes the topological defect is defined in the range: $0<\,\beta\,<\,1$. Therefore, we can neglect the terms proportional to $\beta^{2}$ and write 
\begin{eqnarray}
\lambda\,R+2q^{2}\left(R+2R''\right)\approx\left(l-\beta\,k\right)^{2}\,R,
\label{1.9a}
\end{eqnarray}
without loss of generality. From now on, let us call $\gamma=\left(l-\beta\,k\right)$. With the approximation given in Eq. (\ref{1.9a}), therefore, Eq. (\ref{1.9}) becomes
\begin{eqnarray}
R''+\frac{1}{x}\,R'-\frac{\gamma^{2}}{x^{2}}\,R+R=0.
\label{1.9b}
\end{eqnarray}

Hence, Eq. (\ref{1.9b}) corresponds to the Bessel differential equation \cite{arf,abra}. Since we are dealing with the cylindrical symmetry, hence, we need a regular solution at the origin. Note that when $r=0\Rightarrow x=0$. Therefore, the solution to Eq. (\ref{1.9b}) is given by
\begin{eqnarray}
R\left(x\right)=A\,J_{\left|\gamma\right|}\left(x\right),
\label{1.10}
\end{eqnarray} 
where $J_{\left|\gamma\right|}\left(x\right)$ is the Bessel function of first kind \cite{abra,arf} and $A$ is a constant.

Now, we are able to analyse the confinement of the spinless quantum particle to a hard-wall confining potential in the elastic medium that contains the screw dislocation (\ref{1.1}). This confinement is described by imposing that the wave function vanishes at a fixed value of $x$, i.e., it must vanish when $x\rightarrow x_{0}=\beta\sqrt{2m\mathcal{E}-k^{2}}\,\cosh y_{0}$ ($r_{0}$ is fixed): 
\begin{eqnarray}
R\left(x_{0}\right)=0.
\label{1.11}
\end{eqnarray}
Therefore, by substituting Eq. (\ref{1.10}) into Eq. (\ref{1.11}), we have that the boundary condition (\ref{1.11}) is satisfied when
\begin{eqnarray}
\beta\sqrt{2m\mathcal{E}-k^{2}}\,\cosh y_{0}=\Theta_{n,\,\gamma}\,,
\label{1.12}
\end{eqnarray}
where $\Theta_{n,\,\gamma}$ is the $n$th zero of the $\gamma$th the Bessel function \cite{griff}. With the relation (\ref{1.12}), we obtain the energy levels of the system:
\begin{eqnarray}
\mathcal{E}_{n,\,l,\,k}=\frac{\Theta_{n,\,\gamma}^{2}}{2m\,\beta^{2}\,\cosh^{2}y_{0}}+\frac{k^{2}}{2m}.  
\label{1.13}
\end{eqnarray}
By using Eq. (\ref{1.4}), we can rewrite Eq. (\ref{1.13}) in the form:
\begin{eqnarray}
\mathcal{E}_{n,\,l,\,k}=\frac{\Theta_{n,\,\gamma}^{2}}{2m\,r_{0}^{2}}+\frac{k^{2}}{2m}. 
\label{1.13a}
\end{eqnarray}

Hence, Eq. (\ref{1.13}) or (\ref{1.13a}) yields the allowed energies of the system. Note that the parameters $n$ and $\gamma=\left(l-\beta\,k\right)$ that determine $\Theta_{n,\,\gamma}$ play the role of the quantum number associated with the radial modes and the angular momentum (effective angular momentum), respectively. Therefore, we have that there exists the influence of the topological defect on the energy levels (\ref{1.3}). Besides, this influence of the topological defect is made by a shift in the angular momentum quantum number even though no interaction between the quantum particle and the topological defect exists. This gives rise to an analogue of the Aharonov-Bohm effect for bound states \cite{pesk,fur6,fur2}.

The Aharonov-Bohm effect for bound states that arises from the presence of linear topological defects in an elastic medium has been investigated in several quantum systems \cite{fur6,fur3,dantas1,1,shell,bf3,fur,fur2,fur7,fil,bf,mb,valdir}. In particular, by considering the distortion of a circular curve into a vertical spiral (a screw dislocation), it has been shown in Ref. \cite{fur6} that the energy levels of a spinless quantum particle confined to a cylindrical wire depend on an effective angular momentum. This effective angular momentum corresponds to a shift in the angular momentum quantum. In the present case, the parameter $\gamma=\left(l-\beta\,k\right)$ plays the role of this effective angular momentum. Note that the shift in the angular momentum quantum number is produced by the term $\beta\,k$.

Let us go further with this discussion about the Aharonov-Bohm effect for bound states by analysing a particular case of Eq. (\ref{1.13}). Let us assume that $x_{0}\gg1$. This particular case where $x_{0}\gg1$, when $\left|\gamma\right|$ is fixed, permits us to write the Bessel function as \cite{arf,abra,valdir}:
\begin{eqnarray}
J_{\left|\gamma\right|}\left(x_{0}\right)\rightarrow\sqrt{\frac{2}{\pi\,x_{0}}}\,\cos\left(x_{0}-\frac{\left|\gamma\right|\,\pi}{2}-\frac{\pi}{4}\right).
\label{1.14}
\end{eqnarray}

Therefore, by substituting (\ref{1.14}) into (\ref{1.11}), we obtain
\begin{eqnarray}
\mathcal{E}_{n,\,l,\,k}\approx\frac{\pi^{2}}{2m\beta^{2}\cosh^{2}y_{0}}\left[n+\frac{1}{2}\left|l-\beta\,k\right|+\frac{3}{4}\right]^{2}+\frac{k^{2}}{2m}.
\label{1.15}
\end{eqnarray}
By using Eq. (\ref{1.4}), we can rewrite Eq. (\ref{1.15}) in the form:
\begin{eqnarray}
\mathcal{E}_{n,\,l,\,k}\approx\frac{\pi^{2}}{2m\,r_{0}^{2}}\left[n+\frac{1}{2}\left|l-\beta\,k\right|+\frac{3}{4}\right]^{2}+\frac{k^{2}}{2m},
\label{1.16}
\end{eqnarray}
where $n=0,1,2,\ldots$ is the quantum number related to the radial modes and $l=0,\pm1,\pm2,\ldots$ is the angular momentum quantum number. Therefore, we have obtained a discrete spectrum of energy when the quantum particle is confined to a hard-wall confining potential. Observe the  presence of an effective angular momentum given by $\gamma=\left(l-\beta\,k\right)$ in the energy levels (\ref{1.16}). It stems from the effects of the topology of the distortion of a vertical line into a vertical spiral, i.e., the screw dislocation given in Eq. (\ref{1.1}). Since there is no interaction between the quantum particle and the topological defect, hence, there is the influence of the topology of the defect on the energy levels. This corresponds to an Aharonov-Bohm-type effect. Moreover, by taking $\beta=0$, we recover the energy levels for a spinless quantum particle confined to a cylindrical box.

\section{Effects of rotation}

Recently, effects of rotation on nonrelativistic quantum systems have been reported in the literature by considering a rotating frame with a constant angular velocity given by $\vec{\Omega}=\Omega\,\hat{z}$ \cite{landau3,landau4,dantas1,anan,r13,fb2,fb3,ob}. The time-independent Schr\"odinger equation in this rotating frame is written as \cite{dantas1}: 
\begin{eqnarray}
\hat{H}_{0}\,\Psi-\vec{\Omega}\cdot\hat{L}\,\Psi=\mathcal{E}\Psi,
\label{2.1}
\end{eqnarray}
where $\hat{H}_{0}$ is the Hamiltonian operator of a particle system in the absence of rotation and $\hat{L}$ is the angular momentum operator. An interesting point that has been raised in Refs. \cite{valdir,ang1,ang2,dantas1} is the influence of torsion on the $z$-component of the angular momentum $\hat{L}_{z}$. It is shown that the presence of torsion can modify the $z$-component of the angular momentum by yielding additional contributions to this operator. In particular, in Refs. \cite{valdir,dantas1} is shown that the topology of a screw dislocation that corresponds to the distortion of a circular curve into a vertical spiral transforms the operator $\hat{L}_{z}=-i\frac{\partial}{\partial \varphi}$ into the effective operator $\hat{L}_{z}^{\mathrm{eff}}=-i\left(\frac{\partial}{\partial \varphi}-\chi\frac{\partial}{\partial z}\right)$, where $\chi$ is the parameter that characterizes the distortion of a circular curve into a vertical spiral. By analysing Eq. (\ref{1.2}), we have that the topology of the distortion of a vertical line into a vertical spiral (\ref{1.1}) also yields a change in the $z$-component of the angular momentum. From Eq. (\ref{1.2}), we have 
\begin{eqnarray}
 \hat{L}_{z}^{\mathrm{eff}}=-i\left[\frac{\partial}{\partial \varphi}-\beta\frac{\partial}{\partial z}\right].
\label{2.1a}
\end{eqnarray}
Therefore, the angular momentum in Eq. (\ref{2.1}) is defined in terms of the effective operator (\ref{2.1a}). Besides, since $\hat{H}_{0}$ is determined by Eqs. (\ref{1.1a}) and (\ref{1.2}), then, the time-independent Schr\"odinger equation (\ref{2.1}) becomes
\begin{eqnarray}
-2m\mathcal{E}\psi&=&\frac{\partial^{2}\psi}{\partial r^{2}}+\frac{r}{\left(r^{2}-\beta^{2}\right)}\,\frac{\partial\psi}{\partial r}+\frac{1}{\left(r^{2}-\beta^{2}\right)}\frac{\partial^{2}\psi}{\partial\varphi^{2}}-\frac{2\beta}{\left(r^{2}-\beta^{2}\right)}\,\frac{\partial^{2}\psi}{\partial\varphi\partial z}\nonumber\\
[-2mm]\label{2.2}\\[-2mm]
&+&\frac{r^{2}}{\left(r^{2}-\beta^{2}\right)}\frac{\partial^{2}\psi}{\partial z^{2}}-i\,2m\Omega\,\left[\frac{\partial}{\partial\varphi}-\beta\frac{\partial}{\partial z}\right]\psi.\nonumber
\end{eqnarray}

Next, by following the steps from Eq. (\ref{1.2}) to Eq. (\ref{1.9}), we obtain
\begin{eqnarray}
R''+\frac{1}{x}\,R'-\frac{1}{x^{2}}\left[\bar{\lambda}\,R+2\bar{q}^{2}\left(R+2R''\right)\right]+R=0,
\label{2.3}
\end{eqnarray}
where we have defined the parameters:
\begin{eqnarray}
\bar{\lambda}&=&\gamma^{2}+\frac{\beta^{2}}{2}\left[2m\left(\mathcal{E}+\Omega\,\gamma\right)-k^{2}\right];\nonumber\\
[-2mm]\label{2.4}\\[-2mm]
\bar{q}^{2}&=&\frac{\beta^{2}}{4}\left[2m\left(\mathcal{E}+\Omega\,\gamma\right)-k^{2}\right],\nonumber
\end{eqnarray}
where $\gamma=\left(l-\beta\,k\right)$. Note that we have replaced $\lambda$ with $\bar{\lambda}$ and $q^{2}$ with $\bar{q}^{2}$ in order to obtain the second-order differential equation (\ref{2.3}).

Since $0<\,\beta\,<\,1$, thus, let us also neglect the terms proportional to $\beta^{2}$. In this way, without loss of generality, we can also perform the approximation: 
\begin{eqnarray}
\bar{\lambda}\,R+2\bar{q}^{2}\left(R+2R''\right)\approx\gamma^{2}\,R.
\label{2.5}
\end{eqnarray}
Therefore, Eq. (\ref{2.3}) becomes
\begin{eqnarray}
R''+\frac{1}{x}\,R'-\frac{\gamma^{2}}{x^{2}}\,R+R=0,
\label{2.6}
\end{eqnarray}
which is also the Bessel differential equation \cite{arf,abra}. By following the steps from Eq. (\ref{1.10}) to Eq. (\ref{1.13}), we obtain
\begin{eqnarray}
\mathcal{E}_{n,\,l,\,k}=\frac{\Theta_{n,\,\gamma}^{2}}{2m\,\beta^{2}\,\cosh^{2}y_{0}}+\frac{k^{2}}{2m}-\Omega\,\gamma.  
\label{2.7}
\end{eqnarray}
By using the relation (\ref{1.4}), we rewrite Eq. (\ref{2.7}) in the form:
\begin{eqnarray}
\mathcal{E}_{n,\,l,\,k}=\frac{\Theta_{n,\,\gamma}^{2}}{2m\,r_{0}^{2}}+\frac{k^{2}}{2m}-\Omega\,\gamma.  
\label{2.7a}
\end{eqnarray}

Hence, we have obtained in Eq. (\ref{2.7a}) the spectrum of energy when the spinless particle is confined to a hard-wall confining potential in the rotating reference frame. Again, the parameter $\Theta_{n,\,\gamma}$ is determined by the effective angular momentum $\gamma=\left(l-\beta\,k\right)$ and the quantum number associated with the radial modes $n$. This effective angular momentum stems from the topological effects of the distortion of a vertical line into a vertical spiral. In this sense, there is an analogue of the Aharonov-Bohm effect for bound states. The contribution to the energy levels (\ref{2.7a}) that arises from the effects of rotation is the coupling between the effective angular momentum quantum number $\gamma$ and the angular velocity $\Omega$, which is an analogue of the Page-Werner {\it et al} term \cite{r1,r2,r3}. Note that by taking $\Omega\rightarrow0$, we recover the energy levels (\ref{1.13}).

We can also go further by assuming that $x_{0}\gg1$. This gives us a particular case of Eq. (\ref{2.7a}). We can work with this particular case by following the steps from Eq. (\ref{1.14}) to Eq. (\ref{1.15}). Then, we obtain
\begin{eqnarray}
\mathcal{E}_{n,\,l,\,k}\approx\frac{\pi^{2}}{2m\beta^{2}\cosh^{2}y_{0}}\left[n+\frac{1}{2}\left|l-\beta\,k\right|+\frac{3}{4}\right]^{2}+\frac{k^{2}}{2m}-\Omega\left[l-\beta\,k\right].
\label{2.8}
\end{eqnarray}
With the relation (\ref{1.4}), then, Eq. (\ref{2.8}) becomes
\begin{eqnarray}
\mathcal{E}_{n,\,l,\,k}\approx\frac{\pi^{2}}{2m\,r_{0}^{2}}\left[n+\frac{1}{2}\left|l-\beta\,k\right|+\frac{3}{4}\right]^{2}+\frac{k^{2}}{2m}-\Omega\left[l-\beta\,k\right],
\label{2.9}
\end{eqnarray}
where $n=0,1,2,\ldots$ and $l=0,\pm1,\pm2,\ldots$ are also the quantum numbers associated with the radial modes and the angular momentum, respectively. Therefore, in this particular case, we have obtained a discrete spectrum of energy. The presence of an effective angular momentum given by $\gamma=\left(l-\beta\,k\right)$ in the energy levels (\ref{2.8}) shows that there exists an analogue of the Aharonov-Bohm effect for bound states. Note that effective angular momentum given by $\gamma=\left(l-\beta\,k\right)$ is determined by the effects of the topology of the defect, i.e., there is no contribution of the effects of rotation. In this sense, the Aharonov-Bohm-type effect arises from only the effects of the topology of the defect. On the other hand, the contribution to the energy levels (\ref{2.8}) associated with the effects of rotation is the coupling between the effective angular momentum quantum number $\gamma$ and the angular velocity $\Omega$, i.e., it is the analogue of the Page-Werner {\it et al} term \cite{r1,r2,r3}. Note that by taking $\Omega\rightarrow0$, we recover the energy levels (\ref{1.15}). Moreover, by taking $\beta=0$ and $\Omega\neq0$, we obtain the allowed energies for a spinless quantum particle confined to a cylindrical box under the effects of rotation.

\section{Conclusions}

We have investigated the quantum effects of the distortion of a vertical line into a vertical spiral on a spinless particle confined to a cylindrical box (hard-wall confining potential). By neglecting the terms proportional to $\beta^{2}$, we have obtained the allowed energies of the system. We have seen that the allowed energies are determined by the effective angular momentum quantum number $\gamma=\left(l-\beta\,k\right)$. This additional contribution to the angular momentum quantum number stems from the effects of the topology of this defect even though no interaction between the quantum particle and the topological defect exists. Therefore, it corresponds to an analogue of the Aharonov-Bohm effect for bound states \cite{pesk,fur6,fur2}.

We have also analysed the effects of rotation on this system. We have seen that the allowed energies are also determined by the effective angular momentum given by $\gamma=\left(l-\beta\,k\right)$. Observe that there is no contribution to the effective angular momentum quantum number that stems from the effects of rotation. It is determined by the effects of the topology of the defect. Besides, the presence of this effective angular momentum gives rise to an analogue of the Aharonov-Bohm effect for bound states. On the other hand, the effects of rotation gives rise to the coupling between the effective angular momentum quantum number $\gamma$ and the angular velocity $\Omega$ in the allowed energies. This corresponds to the analogue of the Page-Werner {\it et al} term \cite{r1,r2,r3}. Besides, we have seen that by taking $\Omega\rightarrow0$ and $\beta\neq0$, we recover the allowed energies in the absence of rotation, but in the presence of the topological defect. With $\beta=0$ and $\Omega\neq0$, we have the allowed energies for a spinless quantum particle confined to a cylindrical box under the effects of rotation.

\acknowledgments{The authors would like to thank the Brazilian agencies CNPq and CAPES for financial support.}


\begin{thebibliography}{99}


\bibitem{kleinert} H. Kleinert,  {\it Gauge fields in condensed matter, vol. 2}, (World Scientific, Singapore, 1989).

\bibitem{kat} M. O. Katanaev and I. V. Volovich. Ann. Phys. (NY) {\bf 216}, 1 (1992).

\bibitem{semi} D. L. Dexter and F. Seitz, {\it Effects of Dislocations on Mobilities in Semiconductors}, Phys. Rev. {\bf86}, 964 (1952). DOI: 10.1103/PhysRev.86.964.

\bibitem{semi3} T. Figielski, {\it Dislocations as electrically active centres in semiconductors?alf a century from the discovery}, J. Phys.: Condens. Matter {\bf14}, 12665 (2012). DOI: 10.1088/0953-8984/14/48/301.

\bibitem{semi4} R. Jaszek, {\it Carrier scattering by dislocations in semiconductors}, J. Mater. Sci. Mater. Electron. {\bf12}, 1 (2001). DOI: 10.1023/A:1011228626077

\bibitem{au} E. Aurell, {\it Torsion and electron motion in quantum dots with crystal lattice dislocations}, J. Phys. A: Math. Gen. {\bf32}, 571 (1999). DOI: 10.1088/0305-4470/32/4/004. 

\bibitem{val} K. C. Valanis e V. P. Panoskaltsis, {\it Material metric, connectivity and dislocations in continua}, Acta Mech. {\bf175}, 77 (2005). DOI: 10.1007/s00707-004-0196-9.


\bibitem{put} R. A. Puntigam and H. H. Soleng, {\it Volterra Distortions, Spinning Strings, and Cosmic Defects}, Class. Quantum Grav. {\bf14}, 1129 (1997). DOI: 10.1088/0264-9381/14/5/017. 

\bibitem{fur3} C. Furtado, V. B. Bezerra and F. Moraes, {\it Quantum scattering by a magnetic flux screw dislocation}, Phys. Lett. A {\bf289}, 160 (2001). DOI: 10.1016/S0375-9601(01)00615-6.

\bibitem{fur6} A. L. Silva Netto, C. Chesman and C. Furtado, {\it Influence of topology in a quantum ring}, Phys. Lett. A {\bf372}, 3894 (2008). DOI: 10.1016/j.physleta.2008.02.060.

\bibitem{dantas1} L. Dantas, C. Furtado and A. L. Silva Netto, {\it Quantum ring in a rotating frame in the presence of a topological defect}, Phys. Lett. A {\bf379}, 11 (2015). DOI:10.1016/j.physleta.2014.10.016.

\bibitem{1} N. Soheibi {\it et al}, {\it Screw dislocation and external fields effects on the Kratzer pseudodot}, Eur. Phys. J. B {\bf90}, 212 (2017). DOI: 10.1140/epjb/e2017-80468-9.

\bibitem{shell} C. Filgueiras and E. O. Silva, {\it 2DEG on a cylindrical shell with a screw dislocation}, Phys. Lett. A {\bf379}, 2110 (2015). DOI: 10.1016/j.physleta.2015.06.035.

\bibitem{bf3} K. Bakke and C. Furtado, {\it One-qubit quantum gates associated with topological defects in solids}, Quantum Inf. Process {\bf12}, 119 (2013). DOI: /10.1016/j.physleta.2015.06.035.

\bibitem{fur} G. de A. Marques {\it et al}, {\it Landau levels in the presence of topological defects}, J. Phys. A: Math. Gen. {\bf34}, 5945 (2001). DOI:10.1088/0305-4470/34/30/306.

\bibitem{fur2} C. Furtado and F. Moraes, {\it Landau levels in the presence of a screw dislocation}, Europhys. Lett. {\bf45}, 279 (1999). DOI:10.1209/epl/i1999-00159-8.

\bibitem{fur7} A. L. Silva Netto and C. Furtado, {\it Elastic Landau levels}, J. Phys.: Condens. Matter {\bf20}, 125209 (2008). DOI: 10.1088/0953-8984/20/12/125209.

\bibitem{fil} C. Filgueiras, M. Rojas, G. Aciole and E. O. Silva, {\it Landau quantization, Aharonov-Bohm effect and two-dimensional pseudoharmonic quantum dot around a screw dislocation}, Phys. Lett. A {\bf380}, 3847 (2016). DOI: 10.1016/j.physleta.2016.09.025.

\bibitem{bf} K. Bakke and C. Furtado, {\it Abelian geometric phase due to the presence of an edge dislocation}, Phys. Rev. A {\bf87}, 012130 (2013). DOI: 10.1103/PhysRevA.87.012130.

\bibitem{mb} A. V. D. M. Maia and K. Bakke, {\it Harmonic oscillator in an elastic medium with a spiral dislocation}, Physica B {\bf531}, 213 (2018). DOI: 10.1016/j.physb.2017.12.045

\bibitem{ab} Y. Aharonov and D. Bohm, {\it Significance of Electromagnetic Potentials in the Quantum Theory}, Phys. Rev. {\bf115}, 485 (1959). DOI: 10.1103/PhysRev.115.485

\bibitem{pesk} M. Peshkin and A. Tonomura, \textit{The Aharonov-Bohm Effect} (Springer-Verlag, in: Lecture Notes in Physics, Vol. 340, Berlin, 1989).

\bibitem{valdir} V. B. Bezerra, {\it Global effects due to a chiral cone}, J. Math. Phys. {\bf38}, 2553 (1997). DOI: 10.1063/1.531995

\bibitem{32} J. de S. Carvalho {\it et al}, Eur. Phys. J. C {\bf57}, 817 (2008).

\bibitem{8} J. Carvalho, A. M. de M. Carvalho, E. Cavalcante and C. Furtado, Eur. Phys. J. C {\bf76}, 365 (2016).

\bibitem{31} M. Hosseinpour, H. Hassanabadi and M. de Montigny, Eur. Phys. J. C {\bf79}, 311 (2019).

\bibitem{arf} G. B. Arfken and H. J. Weber, {\it Mathematical Methods for Physicists, sixth edition} (Elsevier Academic Press, New York, 2005).

\bibitem{abra} M. Abramowitz and I. A. Stegum, \textit{Handbook of mathematical functions} (Dover Publications Inc., New York, 1965).

\bibitem{mat1} N. W. McLachlan, {\it Theory and applications of Mathieu functions} (Clarendon Press, Oxford, UK, 1947).

\bibitem{mat2} H. J. W. M\"uler-Kirsten, {\it Introduction to quantum mechanics: Schr\"odinger equation and path integral} (Word Scientific, Singapure, 2006). 

\bibitem{griff} D. J. Griffiths, {\it Introduction to quantum mechanics, Second Edition}, (Prentice Hall, 2004).

\bibitem{landau4} L. D. Landau and E. M. Lifshitz, \textit{Mechanics, third edition} (Pergamon Press, Oxford, 1980).

\bibitem{landau3} L. D. Landau and E. M. Lifshitz, {\it Statistical Physics - Part 1, 3rd. ed.} (Pergamon Press, New York, 1980).

\bibitem{anan} J. Anandan and J. Suzuki in {\it Relativity in Rotating Frames, Relativistic Physics in Rotating Reference Frame}, Edited by G. Rizzi and M. L. Ruggiero (Kluwer Academic Publishers, Dordrecht, 2004) p 361-369; arXiv:quant-ph/0305081. 

\bibitem{r13} C.-H. Tsai and D. Neilson, {\it New quantum interference effect in rotating systems}, Phys. Rev. A {\bf37}, 619 (1988). DOI: 10.1103/PhysRevA.37.619

\bibitem{fb2} I. C. Fonseca and K. Bakke, {\it Rotating effects on the Landau quantization for an atom with a magnetic quadrupole moment}, J. Chem. Phys. {\bf144}, 014308 (2016). DOI: 10.1063/1.4939525.

\bibitem{fb3} I. C. Fonseca and K. Bakke, {\it Some aspects of the interaction of a magnetic quadrupole moment with an electric field in a rotating frame}, J. Math. Phys. {\bf58}, 102103 (2017). DOI: 10.1063/1.5001564.

\bibitem{ob} A. B. Oliveira and K. Bakke, {\it Effects on a Landau-type system for a neutral particle with no permanent electric dipole moment subject to the
Kratzer potential in a rotating frame}, Proc. R. Soc. A {\bf472}, 20150858 (2016). DOI: 10.1098/rspa.2015.0858.


\bibitem{ang1} Y. M. Cho, D. H. Park, and G. G. Han, {\it Gravitational anyon}, Phys. Rev. D {\bf43}, 1421 (1991). DOI: 10.1103/PhysRevD.43.1421.

\bibitem{ang2} B. Jensen and J. Kucera, {\it On a gravitational Aharonov-Bohm effect}, J. Math. Phys. {\bf34}, 4975 (1993). DOI: 10.1063/1.530335.

\bibitem{r1} L. A. Page, {\it Effect of Earth's Rotation in Neutron Interferometry}, Phys. Rev. Lett. \textbf{35}, 543 (1975). DOI: 10.1103/PhysRevLett.35.543.

\bibitem{r2} S. A. Werner, J.-L. Staudenmann and R. Colella, {\it Effect of Earth's Rotation on the Quantum Mechanical Phase of the Neutron}, Phys. Rev. Lett. \textbf{42}, 1103 (1979). DOI: 10.1103/PhysRevLett.42.1103.

\bibitem{r3} F. W. Hehl and W.-T. Ni, {\it Inertial effects of a Dirac particle}, Phys. Rev. D \textbf{42}, 2045 (1990). DOI: 10.1103/PhysRevD.42.2045.






\end{thebibliography}
\end{document}